%% file: draft_pub163.tex
\documentclass[aps,prd,tightenlines,superscriptaddress,showpacs,byrevtex]{revtex4}
\usepackage{times}
\usepackage{amsmath}
\usepackage{graphicx}
\usepackage{epstopdf}
\usepackage{color}
\usepackage{multirow}
\usepackage{setspace}
\usepackage{siunitx}

\newcommand{\lum}{\mathcal{L}}

\newcommand{\kk}{K^+ K^-}

\newcommand{\EE}{e^+e^-}

\newcommand{\kks}{K \overline{K}^{*}}
\newcommand{\ksks}{K^{*} \overline{K}^{*}}

\newcommand{\pipi}{\pi^+ \pi^-}
\newcommand{\etap}{\eta^{\prime}}

\newcommand{\psip}{\psi(3686)}
\newcommand{\psipp}{\psi(3770)}
\newcommand{\psp}{\psi(3686)}

\newcommand{\jpsi}{J/\psi}
\newcommand{\etaY}{\eta Y(2175)}
\newcommand{\etapY}{\etap Y(2175)}
\newcommand{\Y}{Y(2175)}
\newcommand{\zs}{Z_{s}}

\newcommand{\incfig}[2]{\includegraphics[width=#1\textwidth]{#2}}

\def\Journal#1#2#3#4{{#1} {\bf #2}, #3 (#4)}

\def\NIMA{Nucl. Instrum. Meth. A}
\def\NPA{Nucl. Phys. A}

\def\PLB{Phys. Lett. B}
\def\PRL{Phys. Rev. Lett.}
\def\PRD{Phys. Rev. D}

\def\CPC{Chin. Phys. C}
\def\EPJC{Eur. Phys. J. C}
\def\HEPNP{HEP \& NP}

\begin{document}
\DeclareGraphicsExtensions{.eps,.png,.ps}

\title{\boldmath
Observation of $\EE\to \etaY$ at center-of-mass energies above 3.7~GeV}

\input{authors_May2016}

\date{\today}

\begin{abstract}

The state $\Y$ is observed in the process $\EE\to
\etaY$ at center-of-mass energies between 3.7 and
4.6~GeV with a statistical significance larger than $10\sigma$
using data collected with the BESIII detector operating at the
BEPCII storage ring.  This is the first observation of the $\Y$ in this process. The mass and width of the $\Y$ are determined to be
($2135\pm 8\pm 9$)~MeV/$c^2$ and ($104\pm 24\pm 12$)~MeV, respectively,
and the production cross section of $\EE\to \etaY\to \eta\phi
f_{0}(980)\to \eta\phi \pipi$ is at a several hundred
femtobarn level. No significant signal for the process $\EE\to \etap \Y$ is
observed and the upper limit on $\sigma(\EE\to \etap
\Y)/\sigma(\EE\to \eta \Y)$ is estimated to be 0.43 at the 90\%
confidence level. We also search for $\psip\to \eta\Y$. No
significant signal is observed, indicating a strong suppression relative to the corresponding $J/\psi$ decay, in violation of the ``12\% rule."

\end{abstract}


\maketitle


\lefthyphenmin=2
\righthyphenmin=2
\uchyph=0

\section{Introduction}

The $\Y$ was first observed in 2006 by the BaBar
collaboration~\cite{babar_1} via the initial-state-radiation (ISR)
process $\EE\to \gamma_{\rm ISR}\phi f_{0}(980)$ with a mass of
($2175\pm 10\pm 15$)~MeV/$c^2$ and a width of ($58\pm 16\pm 20$)~MeV. It
was subsequently confirmed by the Belle collaboration in the same
process~\cite{belle} and by the BESII and BESIII
collaborations~\cite{bes2_y,bes3_y} in $\jpsi$ hadronic decays. The BaBar
collaboration updated their analysis in 2012 with improved
statistics~\cite{babar_2}.

Behaving similarly to the $Y(4260)$ in the charm sector and
the $\Upsilon(10860)$ in the bottom sector, the $\Y$ is regarded as
a candidate for a tetraquark state~\cite{tetra_1,tetra_2}, a
strangeonium hybrid state~\cite{hybrid}, or a conventional
$s\bar{s}$ state~\cite{conven_1,conven_2}. The quark
model~\cite{quark_1,quark_2,quark_3} predicts two conventional $s\bar{s}$
states near 2175~MeV/$c^2$, 3$^{3}S_1$ and 2$^3D_1$, but both of them
are significantly broader than the $\Y$, which makes the
$\Y$ more mysterious.

Despite all previous experimental and theoretical effort, our knowledge
of the $\Y$ is still very poor.
Its observed production
mechanisms are so far limited to direct $\EE$ annihilation and $\jpsi\to
\etaY$ decay.  Furthermore,
there are inconsistencies in previous mass and width measurements~\cite{belle,babar_2,bes3_y}.

Since the process $\jpsi\to \etaY$ has been observed~\cite{bes2_y,bes3_y}, it is
natural to expect the production of $\etaY$ in $\psip$ decays as
well as in direct $\EE$ annihilation in the non-resonant energy
region. The $\eta$ is a mixture of the pseudoscalar SU(3) octet
and singlet states, therefore the other mixture partner, $\etap$,
is also expected to accompany the production of the $\Y$ when
 the center-of-mass (c.m.) energy ($\sqrt{s}$) of $\EE$ annihilation
 is above the production threshold.
 BESIII has accumulated the world's largest data samples
at the $\psip$ peak and at higher energies up to 4.6~GeV,
which gives us a good opportunity to search for these processes.

\begin{table*}[htbp]
 \centering
\caption{Summary of the data samples and the cross section
measurements of $\EE\to \eta\Y\to\eta\phi f_{0}(980)\to\eta\phi\pipi$. Here $\sqrt{s}$ is the c.m.\
energy, $\mathcal{L}_{\rm int}$ is the integrated luminosity,
$N^{\rm obs}$ is the number of observed signal events from the
simultaneous fit, $(1+\delta)\cdot\epsilon$ is the product of the ISR
correction factor and efficiency. The correction
factors of vacuum polarization, $1+\delta^{\rm vac}$, are listed
except for $\sqrt{s}=3.686$~GeV since the contribution of vacuum
polarization is included in the parameters of the
$\psi(3686)$. Born cross sections $\sigma^{\rm B}$ are listed
with statistical (first) and systematic (second) uncertainties.
The last column is the corresponding statistical significance for
each data sample.} \label{tab:cross_section}
\begin{tabular}{c S S[separate-uncertainty] c c c c}
 \hline\hline
 $\sqrt{s}$~(GeV) & {$\mathcal{L}_{\rm int}$~(pb$^{-1}$)} &
 {$N^{\rm obs}$}    & $(1+\delta)\cdot\epsilon$  &
 $1+\delta^{\rm vac}$ & $\sigma^{\rm B}$~(pb) & Significance \\
 \hline
3.686 & 666  & 19.0\pm9.0 & 0.0861  & -     & 1.72 $\pm$ 0.82 $\pm$ 1.00 & $1.5\sigma$\\
3.773 & 2917 & 47.4\pm9.1 & 0.0865  & 1.057 & 0.93 $\pm$ 0.18 $\pm$ 0.15 & $6.2\sigma$\\
4.008 & 482  & 3.8\pm2.6  & 0.0976  & 1.044 & 0.40 $\pm$ 0.27 $\pm$ 0.34 & $1.0\sigma$\\
4.226 & 1092 & 12.3\pm4.1 & 0.1052  & 1.056 & 0.53 $\pm$ 0.17 $\pm$ 0.05 & $3.8\sigma$\\
4.258 & 826  & 11.6\pm3.7 & 0.1067  & 1.054 & 0.65 $\pm$ 0.21 $\pm$ 0.08 & $4.2\sigma$\\
4.358 & 540  & 6.4\pm2.7  & 0.1113  & 1.051 & 0.53 $\pm$ 0.22 $\pm$ 0.07 & $2.9\sigma$\\
4.416 & 1029 & 10.8\pm4.1 & 0.1135  & 1.053 & 0.46 $\pm$ 0.17 $\pm$ 0.21 & $3.2\sigma$\\
4.600 & 567  & 2.7\pm1.9  & 0.1164  & 1.055 & 0.20 $\pm$ 0.14 $\pm$ 0.02 & $1.5\sigma$\\
\hline
\end{tabular}
\end{table*}

Recently, several charged quarkonium-like states
$Z_c$~\cite{zc_1,zc_2,zc_3,zc_4} and $Z_b$~\cite{zb} have been observed
through decays of the $Y(4260)$, $\Upsilon(10860)$ or other
charmonium-like or bottomonium-like states. One may expect similar
charged strangeonium-like states produced in $\Y\to \phi\pipi$ decays,
considering the similarity of the $\Y$, $Y(4260)$, and
$\Upsilon(10860)$. Ref.~\cite{zs} predicts the existence of a
sharp peaking structure ($Z_{s1}$) close to the $\kks$ threshold and
a broad structure ($Z_{s2}$) close to the $\ksks$ threshold in the
$\pi\phi$ mass spectrum. These can be searched for using
the decays of the $\Y$ produced in
$\EE\to \etaY$ and $\etapY$.

In this article, we present the first observation of $\EE\to \etaY$ and
measurement of its production cross sections, a search for
$\EE\to \etapY$ and an estimation of the upper limit of the
production rate, and the search for $\psip\to \etaY$ and
determination of the upper limit on the branching fraction
at the
c.m.\ energies~\cite{cmsenergy} from 3.686 to 4.6 GeV, as listed
in Table~\ref{tab:cross_section} with
the corresponding integrated luminosities
$\lum$~\cite{luminosity}.

The remainder of this paper is organized as follows: in
Sec.~\ref{sec:detector}, the BESIII detector and the data samples
are described; in Sec.~\ref{sec:selection}, the event selections
for $\EE\to\eta \Y$ are listed; Section~\ref{sec:crosssection}
presents the determination of the signal yield and the cross section measurement, as
well as the measurement of the resonance parameters of the $\Y$ in
$\EE\to\eta\Y$;
while Secs.~\ref{sec:zs} and~\ref{sec:psip} show the search for
the $Z_s$ and $\psp\to \etaY$.
Section~\ref{sec:etap} shows the search for $\EE\to \etapY$,
Sec.~\ref{sec:sys} lists the estimation of the systematic
uncertainties. A summary of all results is given in Sec.~\ref{sec:summary}.


\section{BESIII detector and data samples}
\label{sec:detector}

The BESIII detector, described in detail in Ref.~\cite{bepc2}, has
a geometrical acceptance of $93\%$ of $4\pi$. A small-cell
helium-based main drift chamber (MDC) provides a charged particle
momentum resolution of $0.5\%$ at $1$~GeV/$c$ in a $1$~T magnetic
field, and supplies energy loss ($dE/dx$) measurements with a
resolution better than $6\%$ for electrons from Bhabha scattering.
The electromagnetic calorimeter (EMC) measures photon energies
with a resolution of $2.5\%$ ($5\%$) at $1.0$~GeV in the barrel
(endcaps). Particle identification (PID) is provided by a
time-of-flight system (TOF) with a time resolution of $80$~ps
($110$~ps) for the barrel (endcaps). The muon system, located in
the iron flux return yoke of the magnet, provides $2$~cm position
resolution and detects muon tracks with momentum greater than
$0.5$~GeV/$c$.

The data used in this analysis are listed in
Table~\ref{tab:cross_section}, where the data sample at
$\sqrt{s}=3.686$~GeV corresponds to the $\psip$ data samples of
$(106.8\pm 0.8)\times 10^6$~\cite{npsip_1} and $(341.1\pm
2.1)\times 10^6$~\cite{npsip_2} events collected in 2009 and 2012,
respectively. The data at other energies were taken during 2009
and 2015.

The {\sc geant}4-based~\cite{geant4} Monte Carlo (MC) simulation
software {\sc boost}~\cite{boost} includes the geometric
description of the BESIII detector and a simulation of the
detector response. It is used to optimize event selection
criteria, estimate backgrounds and evaluate the detection
efficiency. For each energy point, signal MC samples of $\EE\to
\etaY$ with $\Y \to \phi f_{0}(980) \to \phi\pipi$, $\phi \to \kk$
and $\eta \to \gamma \gamma$ are generated, and $\eta \Y$
is generated with an angular distribution of $1+\cos^2\theta$ in the $e^+e^-$ c.m. frame.
For the decays of intermediate states, both the $\Y\to\phi f_0(980)$ and $\eta\to\gamma\gamma$
are generated evenly in phase space, and the $\phi\to K^+K^-$ is generated using a {\sc vss} model
in {\sc evtgen}~\cite{evtgen,besevtgen}. The resonant parameters
of the $\Y$ are taken from the measurement in this analysis, and
the $f_{0}(980)$ is parameterized with the Flatt\'{e}
formula~\cite{flatte}, with parameters determined from the BESII
experiment~\cite{flatte_bes2}. The ISR is
simulated with {\sc kkmc}~\cite{kkmc}, and the
final state radiation (FSR) is handled with {\sc
photos}~\cite{photos}.
The process
   $\EE\to\eta' \Y$ is simulated at each energy point with a similar procedure, and the decay
   $\eta'\to\gamma\pi^{+}\pi^{-}$ is generated as $\eta'\to\gamma\rho^{0}$ with $\rho^{0}\to\pi^+\pi^-$~\cite{cleo_etap}.

For background studies, two inclusive MC samples with integrated
luminosities equivalent to those of data are generated at
$\sqrt{s}=3.686$ and $3.773$~GeV. In these samples the $\psip$ and
$\psipp$ are allowed to decay generically, with the main known
decay channels being generated using \textsc{evtgen}~\cite{evtgen} with
branching fractions set to world average values~\cite{pdg}. The
remaining events associated with charmonium decays are generated
with {\sc lundcharm}~\cite{lundcharm} while continuum hadronic
events are generated with {\sc pythia}~\cite{pythia}.
For the QED events,
$\EE\to\tau^+\tau^-$ is generated with {\sc
kkmc}~\cite{kkmc}, and other events are generated with {\sc babayaga}~\cite{babayaga}.

\section{Event selections}
\label{sec:selection}

For the study of $\EE\to\eta\Y$, we expect four charged particles with zero
net charge and two photons in the final state.

Each charged track is required to have its point of closest
approach to the beamline within 1~cm in the radial direction and
within 10~cm from the interaction point along the beam direction,
and to lie within the polar angle coverage of the MDC,
$|\cos\theta|<0.93$ in the laboratory frame. PID for charged tracks is
based on combining the $dE/dx$ and TOF information. The
confidence levels
${\rm Prob_{PID}}(i)$ are calculated for each charged track for
each particle hypothesis $i=$(pion, kaon, or proton). If ${\rm
Prob_{PID}}(K)>{\rm Prob_{PID}}(\pi)$ and ${\rm Prob_{PID}}(K)$ is
larger than 0.001, the track is regarded as a kaon, otherwise it
is taken as a pion. Two identified kaons with opposite charge are required.

Photons are reconstructed from isolated showers in the EMC which
are at least 10 degrees away from the charged tracks. A good photon
 is required to have an energy of at least 25~MeV in the barrel ($|\cos\theta|<0.80$) or
50~MeV in the end-caps ($0.86<|\cos\theta|<0.92$). In order to
suppress electronic noise and energy deposits unrelated to the
event, the EMC time $t$ of the photon candidate must be in
coincidence with the event start time in the range $0\leq t\leq
700~\rm{ns}$. The $\eta$ candidate is reconstructed using the two most
energetic photons.

A four-constraint (4C) kinematic fit, which constrains the
four-momentum of all particles in the final state to be that of
the initial $\EE$ system, is performed for the $\gamma \gamma
\pipi \kk$ system to get a better resolution and background
suppression. The $\chi^{2}$ of the kinematic fit is required
to be less than 60.

\begin{figure}[htbp]
\begin{center}
\incfig{0.49}{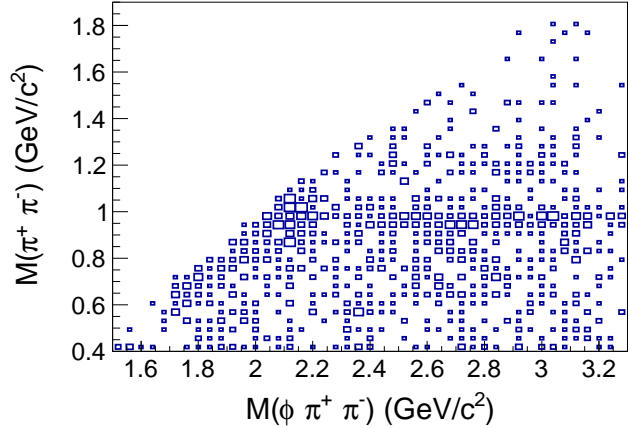} \caption{Scatter plot
of $M(\pipi)$ versus $M(\phi\pipi)$ for the sum of data samples
with $\sqrt{s}>$3.7~GeV.}\label{fig:scatter}
\end{center}
\end{figure}

After all the above selection criteria are applied, we use
mass windows around the $\eta$ and $\phi$, numerically [0.513,
0.578]~GeV/$c^2$ and [1.009, 1.031]~GeV/$c^2$, respectively, to select signal
events. The $\pipi$ system in $\Y\to \phi\pipi$ decays tends to
have $J^{PC}$ = $0^{++}$ and is dominated by $f_{0}(980)$.
Figure~\ref{fig:scatter} shows the scatter plot of $M(\pipi)$
versus $M(\phi\pipi)$ for the sum of the data samples with
$\sqrt{s}>$3.7~GeV. A clear cluster corresponding to the $\Y\to
\phi f_{0}(980)$ events, is visible. Only events in the
mass window of the $f_{0}(980)$ ([0.868, 1.089]~GeV/$c^2$) are used for the
cross section measurement. The mass windows used above are defined
as $[\mu-1.5\cdot W, \mu+1.5\cdot W]$, where $\mu$ and $W$ are the
mean value and full width at half maximum (FWHM), respectively, of
the invariant mass distributions of signal events from the MC
simulation. Analogously, the corresponding sideband regions are
defined as [$\mu-5\cdot W, \mu-2\cdot W$] and [$\mu+2\cdot W,
\mu+5\cdot W$], which are twice as wide as the signal region.

\begin{figure*}[htbp]
\begin{center}
\incfig{0.24}{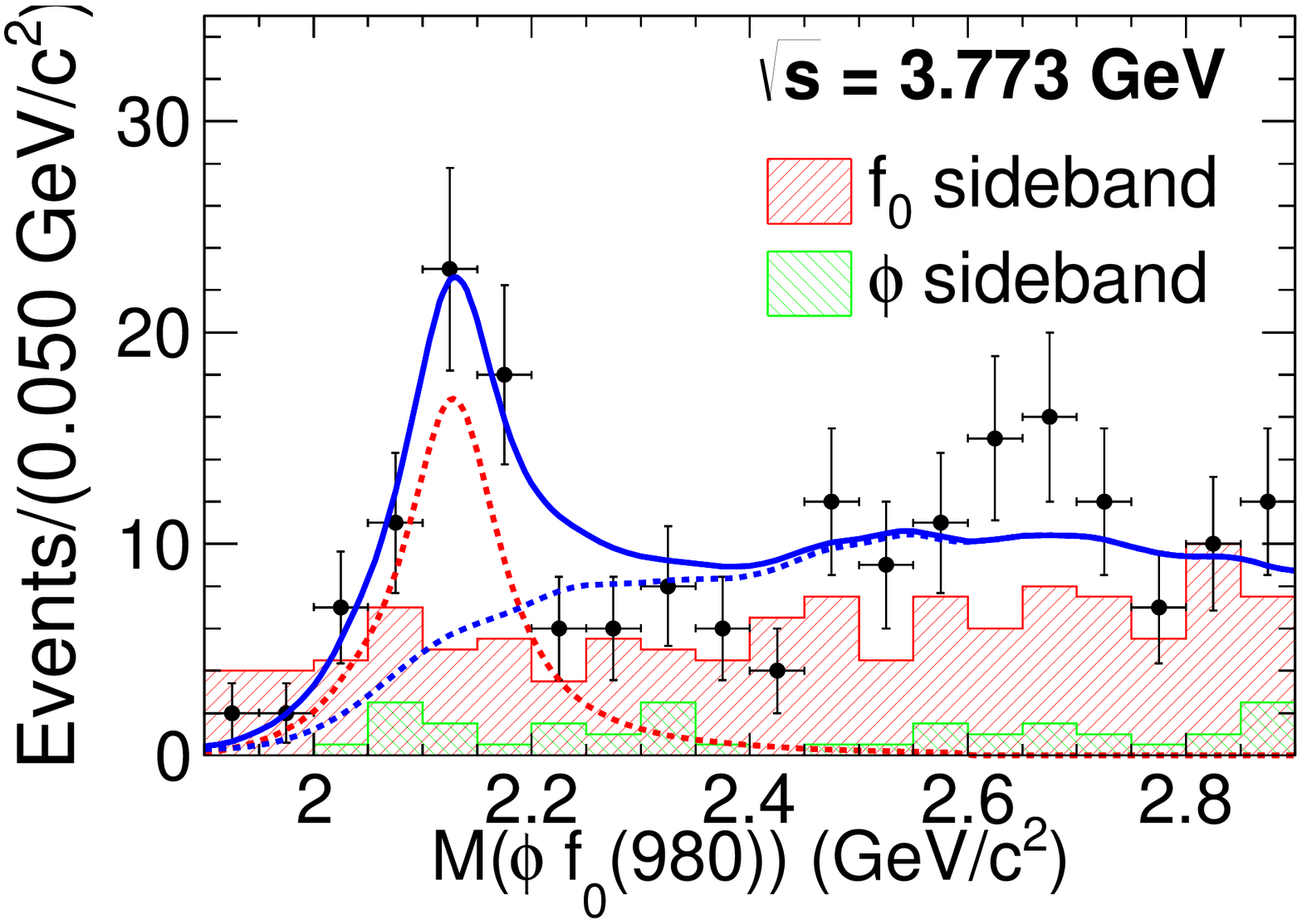} \incfig{0.24}{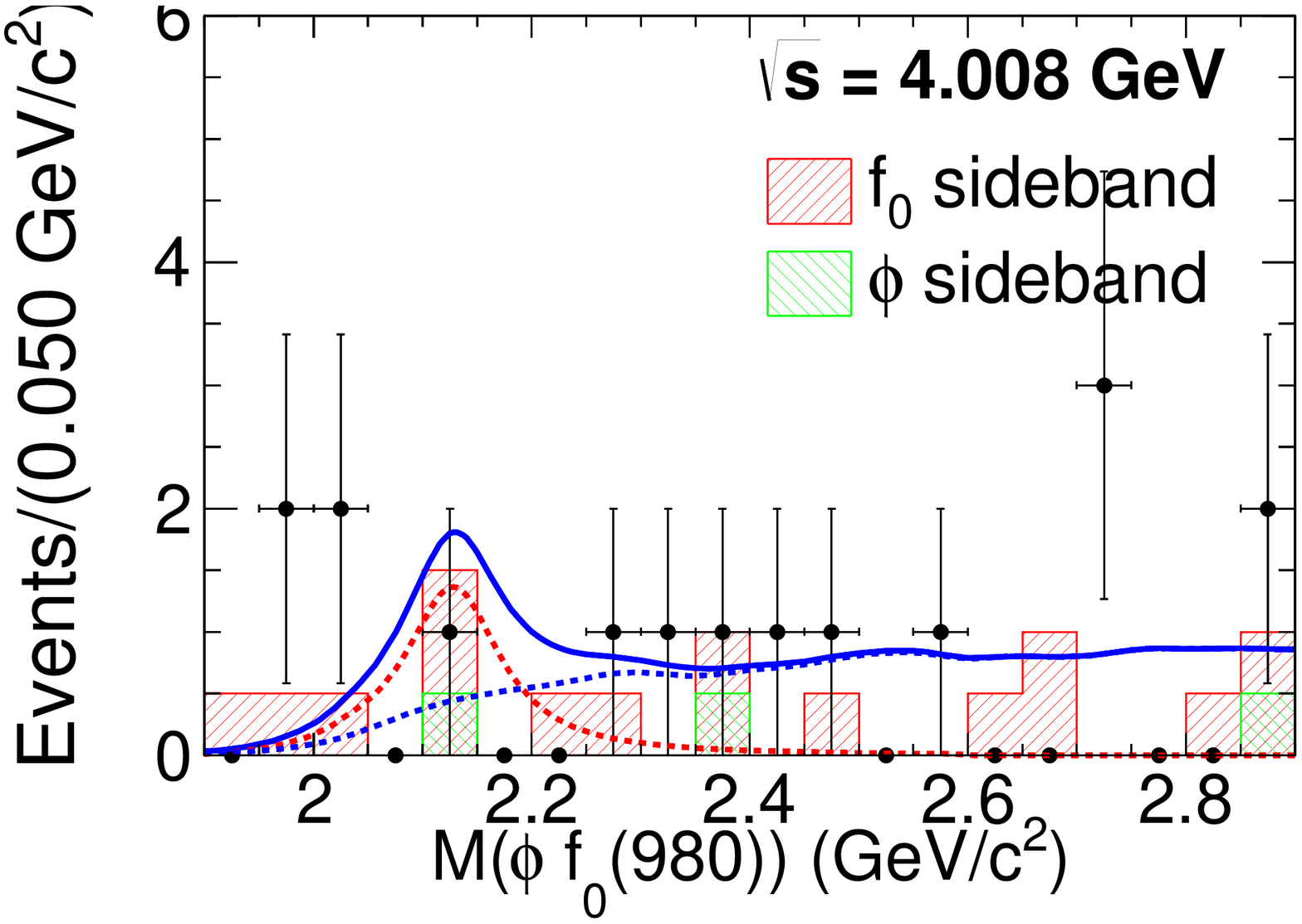}
\incfig{0.24}{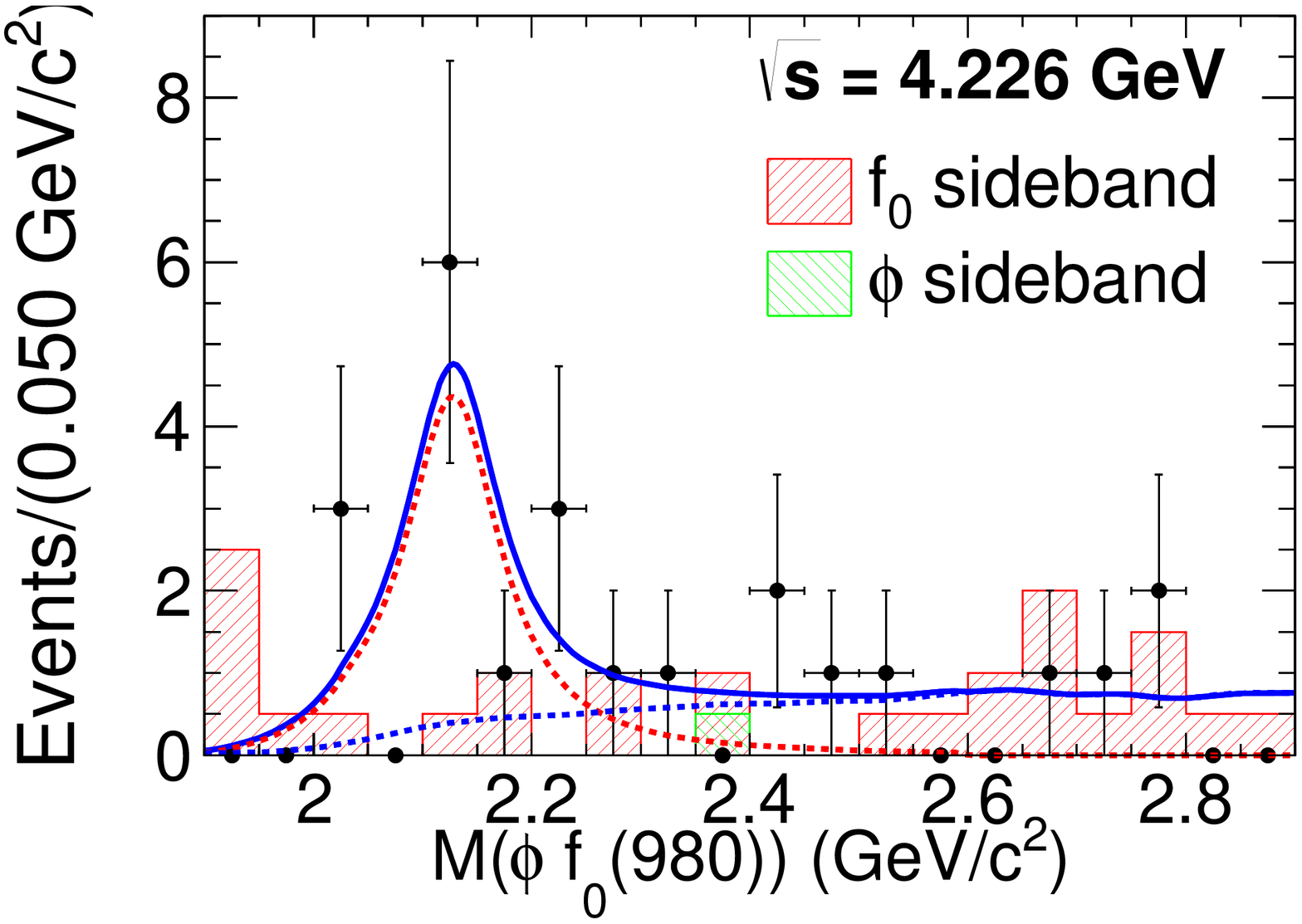} \incfig{0.24}{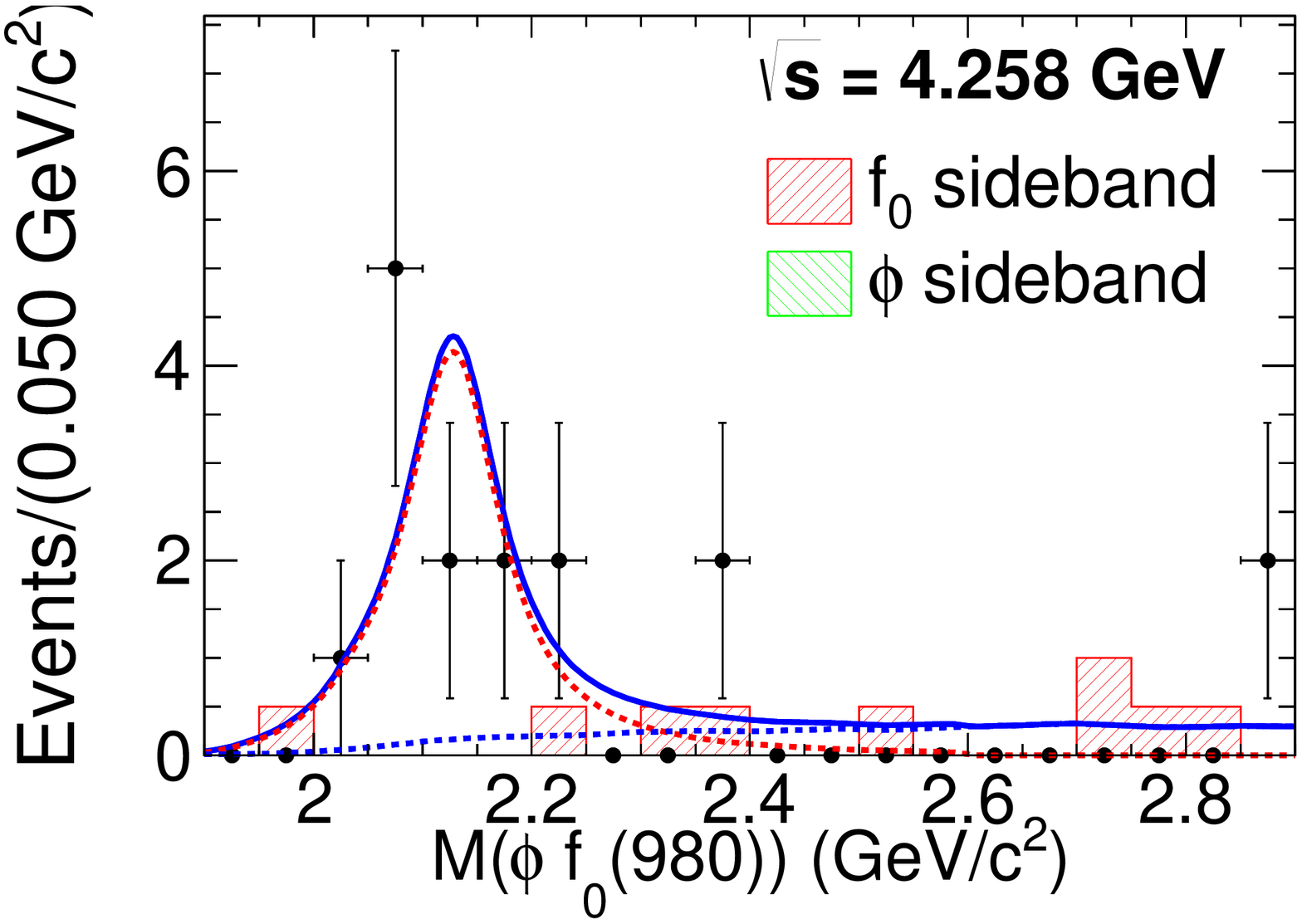} \\
\incfig{0.24}{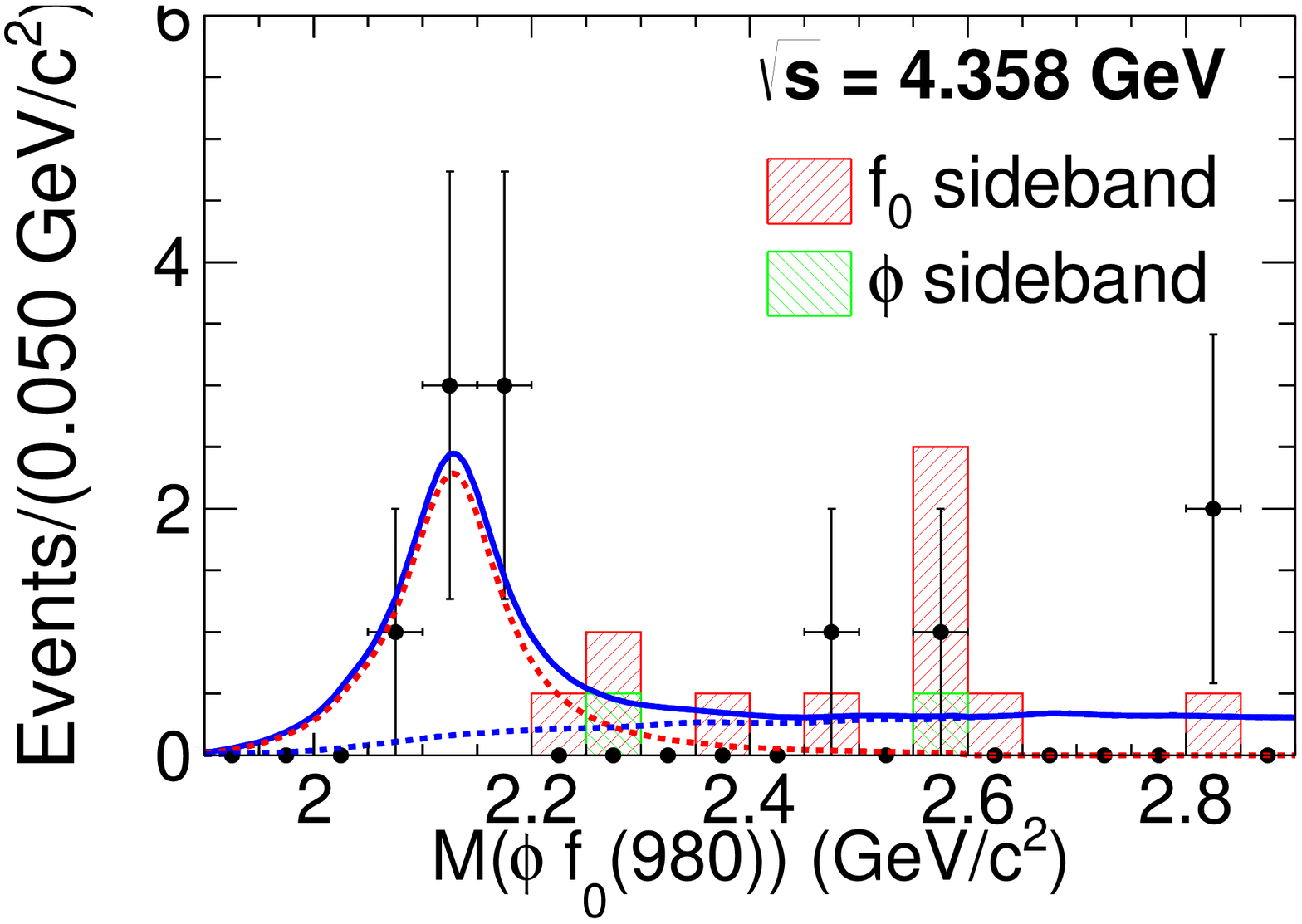} \incfig{0.24}{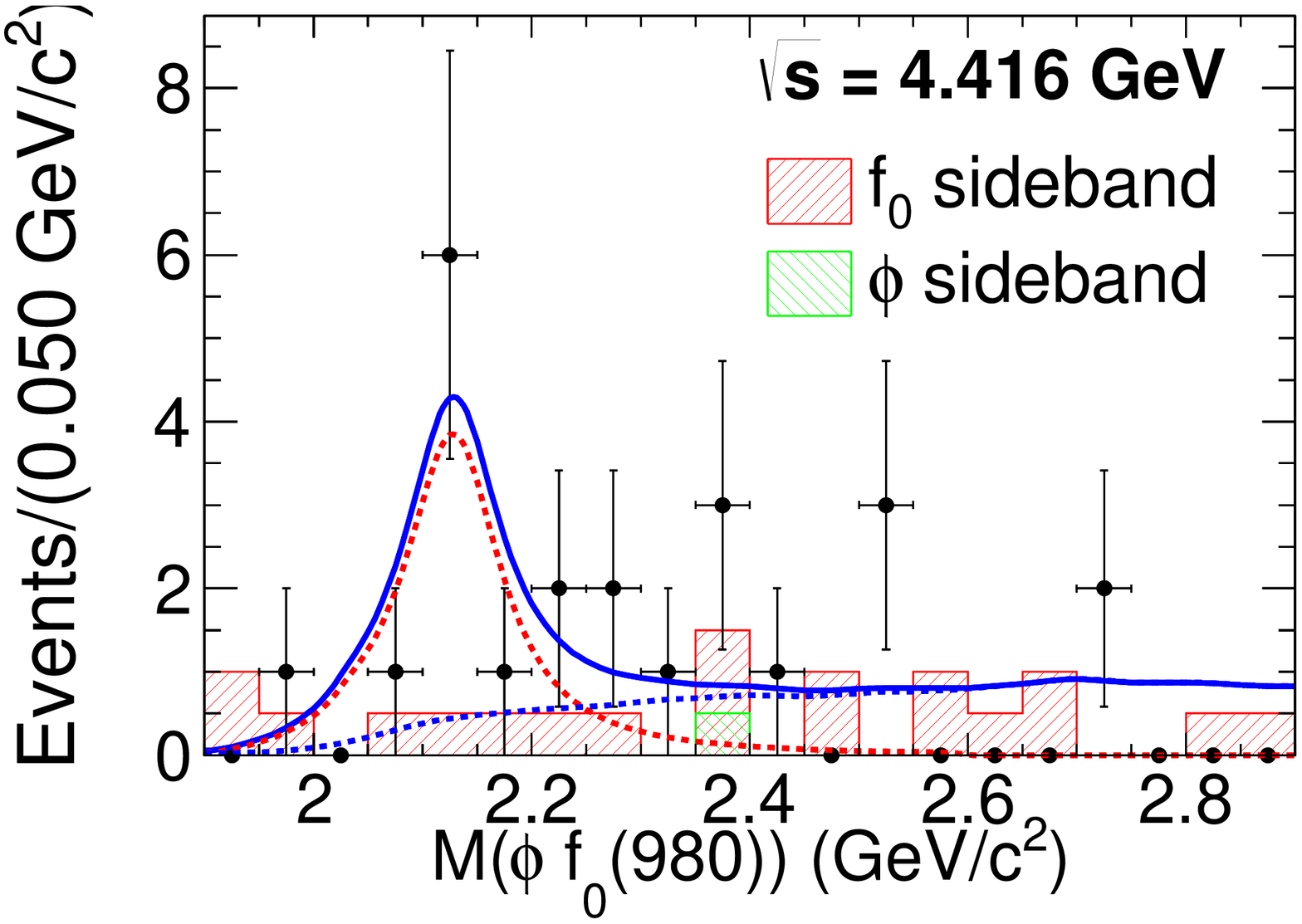}
\incfig{0.24}{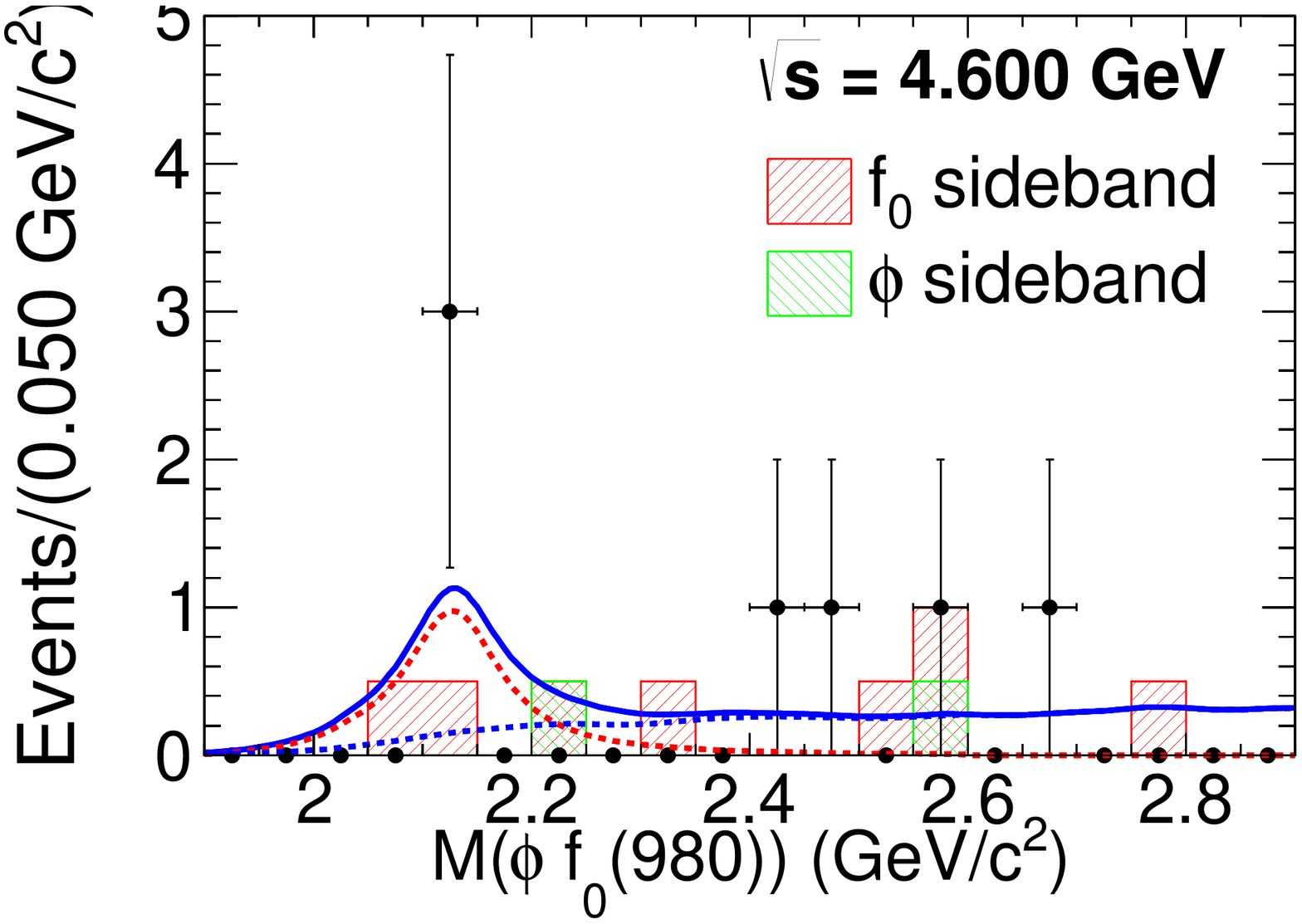} \incfig{0.24}{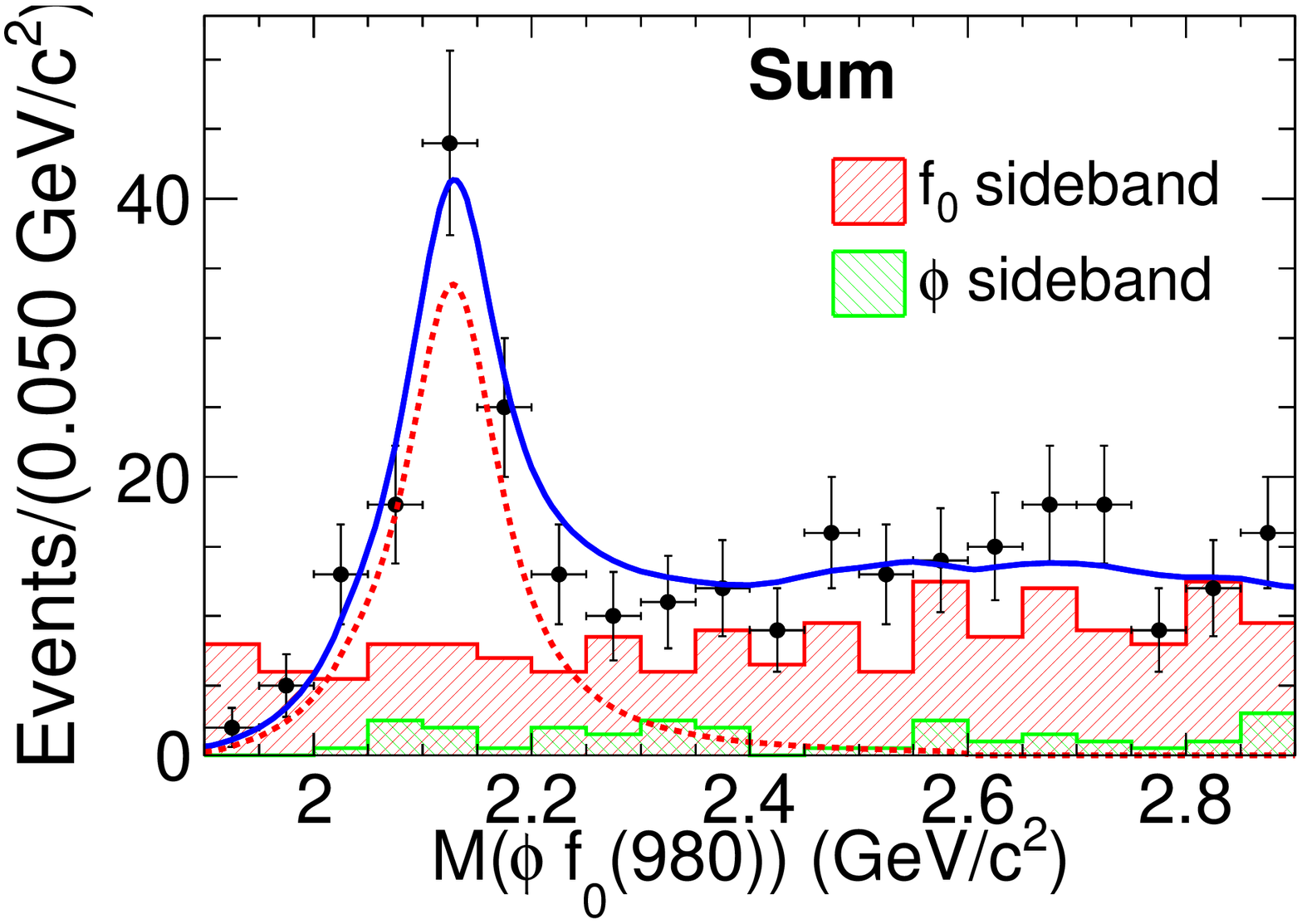}\\
\caption{(Color online) Invariant mass distributions of $\phi f_{0}(980)$ and
the projections of the simultaneous fit (solid curve) at the different c.m.\ energies, as well as the sum of them
(bottom right, marked as ``Sum"). The dots with error bars are
data, the red dotted curves represent the Breit-Wigner functions
for the signal, and the blue dashed curves represent the
background contributions, which are modeled by the MC distribution
for the non-resonant background. The red and green histograms
represent the normalized events from the $f_{0}(980)$ and $\phi$ mass
sideband regions.}\label{fig:fit_simu_1}
\end{center}
\end{figure*}

The invariant mass distribution of $\phi f_{0}(980)$ for
the seven data samples with $\sqrt{s}>$3.7~GeV is shown in
Fig.~\ref{fig:fit_simu_1}, individually. The $\Y$ signal can be observed over
a smooth background level, especially for data sample at 3.773~GeV, where the
integrated luminosity is the largest. The invariant mass
distribution of $\phi f_{0}(980)$ summing over the seven data
samples with $\sqrt{s}>$3.7~GeV is also shown in
Fig.~\ref{fig:fit_simu_1}. We leave the analysis of data at
3.686~GeV to Sec.~\ref{sec:psip} and focus on energy
points with $\sqrt{s}>$3.7~GeV here.

The inclusive MC sample at 3.773~GeV is used to check for possible
backgrounds. No peaking background is found and the main
background is the non-$\Y$ process $e^+e^-\to\eta K^+K^- \pi^+\pi^-$, including
   both the $\eta\phi \pi^+\pi^-$ and $\eta K^+K^- f_0(980)$ processes.
There are almost no other backgrounds around the $\Y$ peak
area. Exclusive MC samples of non-$\Y$ processes are
generated, and the shapes are used to describe the background in
the fit to the invariant mass distributions. Events in the sideband regions
of the $f_0(980)$ and $\phi$ are used to check for the presence of peaking background, and the corresponding distributions
are shown in Fig.~\ref{fig:fit_simu_1}.

\section{Signal yields and Born cross sections}
\label{sec:crosssection}

We use an unbinned maximum likelihood method to fit the $\phi
f_{0}(980)$ invariant mass spectra in order to extract the yields
of signal events and the $\Y$ resonant parameters. A simultaneous fit is applied
to all the data samples with $\sqrt{s}>$3.7~GeV. The same signal
shape is used to describe the signal at different energy points, which is
\begin{equation}
\label{equ:bw} \left(\left |\frac{M\Gamma}{M^{2}-m^{2} -
iM\Gamma}\right |^2 \cdot \frac{\Phi(m)}{\Phi(M)}\cdot
\epsilon(m)\right )\otimes { G(m; 0, \sigma)},
\end{equation}
where $M$ and $\Gamma$ are the mass and width of the $\Y$,
respectively, $G$ is a Gaussian function with a mean fixed to zero
and a free standard deviation $\sigma$ to describe the mass resolution,
$\epsilon(m)$ is the mass-dependent efficiency determined from MC
simulation. $\Phi(m)=(\frac{\mid p \mid}{\sqrt{s}})^3$ is the
two-body phase space factor for a $P$-wave system, where $p$ is
the momentum of $\Y$ in the $\EE$ rest frame. The background shape
is taken from MC simulation of the non-resonant process.

Figure~\ref{fig:fit_simu_1} shows the projections of the simultaneous fit and
their sum. The mass and width of the $\Y$ are determined to be ($2135\pm
8$)~MeV/$c^2$ and ($104\pm 24$)~MeV, respectively, where the uncertainties are
statistical only. The joint statistical significance of the $\Y$
signal is estimated to be larger than $10\sigma$ by comparing the
log-likelihood values with and without the $\Y$
signal included in the fit and considering the change of the number of
degrees of freedom. For each data sample, the statistical
significance is estimated separately by fitting with and without the
$\Y$ signal included while the resonant parameters of the $\Y$ are fixed to the values
of the simultaneous fit. The numbers of signal events and the
statistical significances are listed in
Table~\ref{tab:cross_section}.

The Born cross section of $\EE\to \etaY\to \eta\phi f_0(980)\to
\eta\phi\pipi$ is calculated using
\begin{equation}
\label{equ_2} \sigma^{\rm B} = \frac{\sigma^{\rm obs}}
{(1+\delta)(1+\delta^{\rm vac})} = \frac{N^{\rm obs}}
{\mathcal{L}_{\rm int}\mathcal{B}\epsilon (1+\delta)
(1+\delta^{\rm vac})},
\end{equation}
where $\sigma^{\rm obs}$ is the observed cross section including
the branching fraction $\mathcal{B}(\Y\to \phi f_{0}(980)\to
\phi\pipi)$, $N^{\rm obs}$ is the number of signal events,
$\mathcal{L}_{\rm int}$ is the integrated luminosity,
$\mathcal{B}$ is the product of branching fractions of $\eta\to
\gamma\gamma$ and $\phi\to \kk$, $\epsilon$ is the detection efficiency, and
$(1+\delta)$ is the ISR correction factor, including ISR,
$\EE$ self-energy and initial vertex correction; and the vacuum
polarization factor $(1+\delta^{\rm vac})$, including leptonic and
hadronic contributions, is taken from
Ref.~\cite{vacuum:polarization}.

The vector-pseudoscalar (VP) processes $\EE\to VP$ are predicted to have Born cross sections that vary as $1/s^{n}$~\cite{pqcd_vp} in the absence of
contributions from charmonium(-like) resonances.
 In calculating the ISR correction factors~\cite{qed_radi}, the
Born cross section of $\EE\to \etaY$ from threshold to the c.m.
energy under study is needed as input. We assume the $\etaY$ comes
from a QED process without the contribution from any
 charmonium(-like)
resonances, and the line-shape is parameterized as
\begin{equation}
\label{equ:3} \sigma(s)\propto \frac{1}{s^n}.
\end{equation}
Here $n$ is a parameter describing the energy-dependent form
factor of $\EE\to \etaY$, which is obtained from a fit to the
measured Born cross sections in this analysis. We use an
iterative procedure to measure the Born cross sections and
determine the ISR correction factors together with the efficiencies.

\begin{figure}[htbp]
 \centering \incfig{0.49}{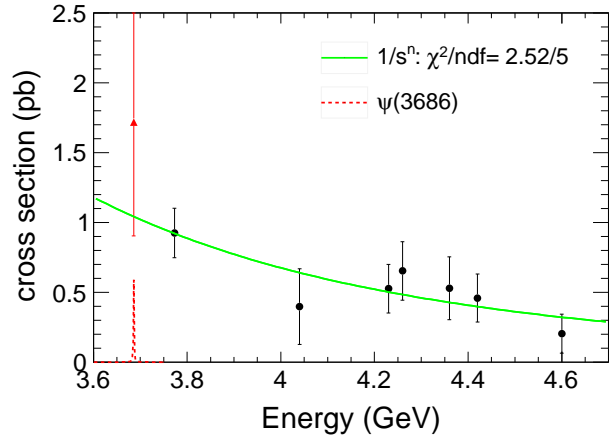}
\caption{
(Color online) Distribution of the Born cross sections for $\EE\to \etaY$ (red triangle for
data at 3.686~GeV and black dots represent the other data
samples). The solid green curve shows the fit result from data samples with $\sqrt{s}>3.7$~GeV
using the shape of Eq.~(\ref{equ:3}) and
the red dashed curve shows the contribution from $\psip$.} \label{fig:fit_cross}
\end{figure}

The resultant Born cross section and all the numbers used in
the calculation are listed in Table~\ref{tab:cross_section} and
shown in Fig.~\ref{fig:fit_cross}. The fit to the final Born cross
sections with Eq.~(\ref{equ:3}) results in $n=2.65\pm 0.86$, as
shown in Fig.~\ref{fig:fit_cross}, and the goodness of fit is $\chi^2/ndf=2.52/5$.

\section{\boldmath Search for $\zs$ states}
\label{sec:zs}

\begin{figure}[htbp]
 \incfig{0.49}{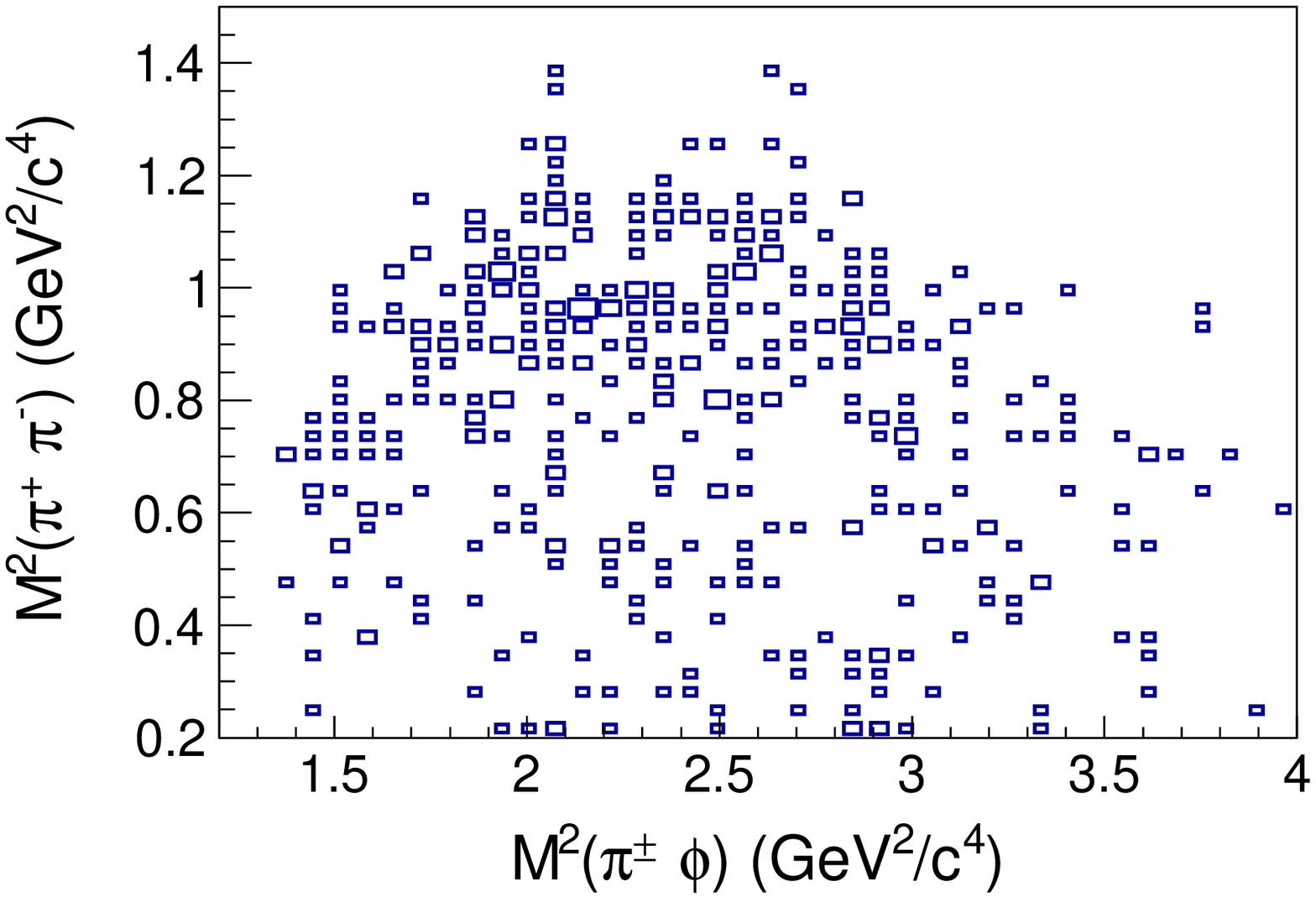}
 \incfig{0.49}{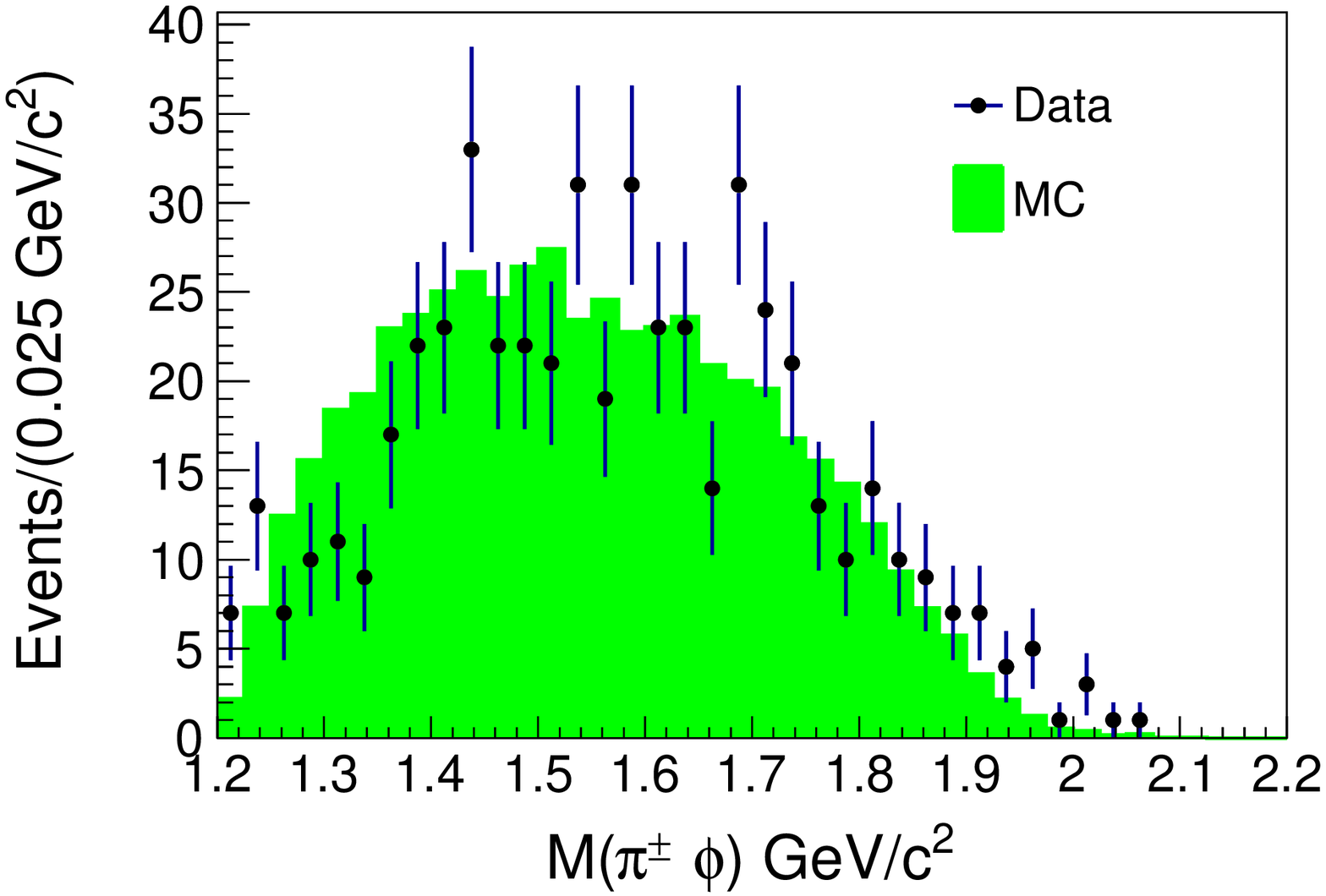}
\caption{Scatter plot (left) and projection on $M(\phi\pi^{\pm})$ (right) of $\Y\to \phi\pipi$ events for the sum of
all data samples above 3.7~GeV (two entries per event).
The green histogram in the right plot shows the same distribution for the normalized
exclusive MC samples of non-$\zs$ process.}
\label{fig:data_zs}
\end{figure}

Since we have observed a distinct $\Y$ signal, possible
charged $\zs$ states in the $\phi\pi^{\pm}$ invariant mass spectrum
can be searched for in the $\Y$ decays. In the cross section
measurement, the candidate events are required to be within the $f_{0}(980)$
mass window to suppress background. This requirement is released
to include the non-$f_0(980)$ decay of $Y$ in the search for
the $\zs$ states. The events in the $\Y$ signal region, [1.989,
2.272]~GeV/$c^2$, are selected and the Dalitz plot of $\Y\to \phi\pipi$
events for the sum of data samples above 3.7~GeV is shown in
Fig.~\ref{fig:data_zs} (left). A clear $f_{0}(980)$ band in the horizental
direction is observed which dominates the $\Y\to \phi\pipi$
decays. Figure~\ref{fig:data_zs} (right) shows the projection on $M(\phi\pi^{\pm})$
for data and MC simulations of the non-$Z_s$ process, which covers all the
energy points and is normalized according
to the luminosity and the fit result in Fig.~\ref{fig:fit_cross}.

From the theoretical calculation~\cite{zs}, which assumes the
$\zs$ states being $\kks$ and $\ksks$ molecular states, the masses of
$\zs$ states are expected at around 1.4 and 1.7~GeV/$c^2$. No significant
vertical bands can be seen at the expected positions. We do not try
to give quantitative results on the $\zs$ production due to the
limited statistics and the not well-defined masses and widths of
these states.

It is worth noting that the $\Y$ signal produced in $\EE\to \etaY$
at c.m. energies above 3.7~GeV has a much lower background level
compared with those in the other two known production processes,
i.e., $\EE$ annihilation around the $\Y$
peak~\cite{babar_1,babar_2,belle} and $J/\psi\to
\etaY$~\cite{bes2_y,bes3_y}, though the signal yield is not comparable to the later two processes at BESIII. With more data accumulated above
3.7~GeV, the $\zs$ states could be searched for with high
sensitivity via $\EE\to \etaY$.

\section{\boldmath Search for $\psi(3686)\to \eta \Y$}
\label{sec:psip}

\begin{figure}[htbp]
\begin{center}
 \incfig{0.49}{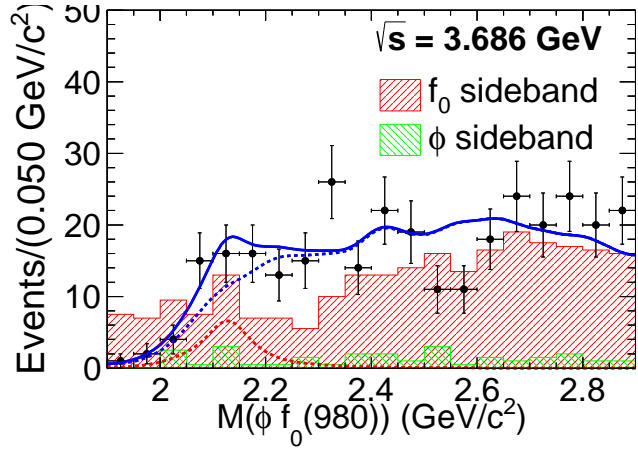}
\caption{(Color online) Invariant mass distribution of $\phi f_{0}(980)$ at
c.m. energy 3.686~GeV and the fit result (solid curve). The dots with
error bars are data, the red dashed curve represents the
Breit-Wigner function for the signal, and the blue dashed curve
represents the background contribution, which is modeled by the MC
simulation for the non-resonant background. The red and green
histograms represent the events from the $f_{0}(980)$ and $\phi$
mass sideband regions.}\label{fig:psp}
\end{center}
\end{figure}

With the same selections as those described in
Sec.~\ref{sec:selection}, the $\phi f_{0}(980)$ invariant mass
distribution at c.m.\ energy 3.686~GeV is shown in
Fig.~\ref{fig:psp}. In contrast to the distributions at $\sqrt{s}>3.7$~GeV, no
significant $\Y$ signal is observed. The background level is
much higher than that at other energies, considering the difference
in integrated luminosities, indicating that
$\psip$ decays are the main background.

The inclusive MC sample at 3.686~GeV is used to check for possible
backgrounds. No peaking background is found and the main
background is the non-$\Y$ process $\psip\to \eta\kk\pipi$ (as well as a
small fraction of $\EE\to \eta\kk\pipi$ through continuum
production), including
   both the $\eta\phi \pi^+\pi^-$ and $\eta K^+K^- f_0(980)$ processes, and there are no
other kinds of background around the $\Y$ peak area. Exclusive
MC samples of non-$\Y$ processes are generated and the shapes
are used to describe the background in the fit to the invariant mass
distribution as in the analysis of data with
higher energy.

The same fit functions for signal and background as in
the fit to the data with higher energy (Sec.~\ref{sec:crosssection}) are used
to determine the signal yield of $\Y$. Since
the signal yield of $\Y$ is very small, we fix the
mass and width of the $\Y$ to the values obtained in the previous fit. The fit returns $19.0\pm 9.0$ events of $\Y$ signal with a
statistical significance of $1.5\sigma$.  The fit curve is shown in
Fig.~\ref{fig:psp}. The Born cross section and all other numbers used
to calculate the Born cross section are listed in
Table~\ref{tab:cross_section} and are shown in Fig.~\ref{fig:fit_cross}.
The cross section does not show any peaking structure, within the large experimental uncertainties.  We therefore assume that the $\Y$ signal is due to continuum production only.

As the process $J/\psi\to \etaY$ has been observed~\cite{bes2_y,bes3_y}, we
expect the production of $\psip\to \etaY$ to occur as well, although there is no
guideline for a prediction of the decay branching fraction.
This branching fraction can be obtained by combining the
measured Born cross sections at
3.686~GeV and those at higher energies which serve to estimate the continuum
cross section at 3.686~GeV.

As described in Sec.~\ref{sec:crosssection}, the obtained Born cross sections for the data samples above 3.7~GeV are fitted
based on an assumption that no charmonium(-like) resonances above
open-charm threshold contribute to this decay. Hence the extrapolation from the fit
result, which is shown as the green curve in
Fig.~\ref{fig:fit_cross}, is used to estimate the contribution
from the QED continuum process at 3.686~GeV. After subtracting the
contribution from QED process (assuming there is no interference
between the resonant and QED processes~\cite{rule12}), we obtain the
product $\sigma(\EE\to \psip)\cdot \mathcal{B}(\psip\to \etaY
\to \eta\phi f_{0}(980)\to \eta \phi \pipi)=(0.68\pm 0.82$)~pb,
with the number of signal events estimated to be $7.5\pm 9.0$,
where the errors are statistical only. The efficiency, 10.7\%, is
obtained from an exclusive MC simulation of $\psip\to \etaY$.
Using the total number of produced $\psi(3686)$
events~\cite{npsip_1,npsip_2}, we obtain
$\mathcal{B}(\psip\to\etaY) \cdot\mathcal{B}(\Y\to \phi
f_{0}(980)\to \phi \pipi)=(0.81\pm 0.97)\times 10^{-6}$, or less
than $2.2\times 10^{-6}$ at the 90\% confidence level (C.L.),
where the systematic uncertainty, which will be detailed later, has been included. The Bayesian method, as described in Ref.~\cite{bes3_y4140}, is used to estimate the
upper limit.

Using $\mathcal{B}(J/\psi\to \etaY)\cdot\mathcal{B}(\Y\to \phi
f_{0}(980)\to \phi\pipi)=(1.20\pm 0.14({\rm stat.})\pm 0.37({\rm
syst.}))\times 10^{-4}$ from previous BESIII's
measurement~\cite{bes3_y}, the ratio of the reduced branching
fractions $\mathcal{B^*}(\psip\to \etaY)/\mathcal{B^*}(J/\psi\to
\etaY)$ is obtained to be $=(0.23\pm 0.29)$\% after considering the
two-body $P$-wave phase space between $\psip$ and $J/\psi$ decays,
 or
less than 0.65\% at the 90\% C.L. after considering the uncertainty of $\mathcal{B}(J/\psi\to \etaY)$.
Here the uncertainty is statistical only. We find that the ratio is
much smaller than the ``12\% rule" expectation and follows the
same suppression pattern as the other two-body VP
final states known as the ``$\rho\pi$ puzzle"~\cite{rule12}.

\section{\boldmath Search for $\EE\to \etapY$}
\label{sec:etap}

For $\EE\to \etapY$, the analysis is
similar to that of $\EE\to\etaY$. The difference occurs in
the reconstruction of the $\eta'$. We use the decay mode $\gamma \pi^{+}
\pi^{-}$ to reconstruct $\etap$, and use the same final state to
reconstruct the $\Y$ as in the $\EE\to\eta\Y$ case. There are four
charged pions and two charged kaons in the final state. To
classify these particles, we first use PID to separate  kaons from
pions, and use a kinematic fit to identify the $\pipi$ from $\etap$
decays. The fit enforces energy-momentum conservation and the
invariant mass of $\gamma \pi^{+} \pi^{-}$ is constrained to the
nominal $\etap$ mass. We loop over all the $\pipi$ combinations,
and the one with the smallest $\chi^2$ is retained. In
order to use the information of the $\etap$ sideband for further
study, the $\etap$ mass constraint is released after the $\pipi$
from $\etap$ decays is identified and the $\chi^{2}$ of the 4C
kinematic fit is required to be less than 60. Mass windows of
$\etap$ ([0.943,0.971] GeV/$c^2$), $f_{0}(980)$, and $\phi$ are used to select signal
events.

\begin{figure}[htbp]
 \incfig{0.49}{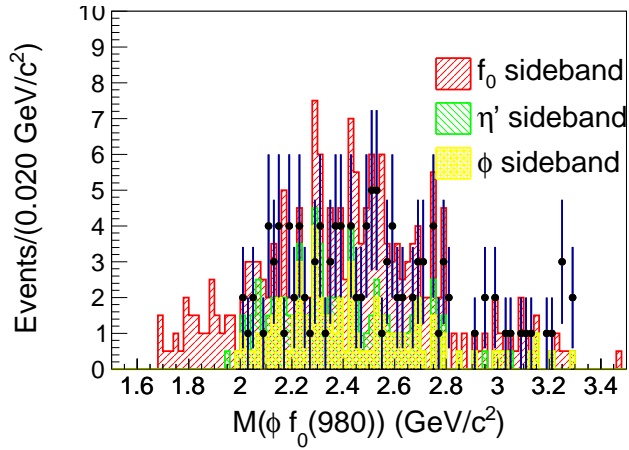}
\caption{(Color online) Distribution of $M(\phi f_{0}(980))$ for the sum of data
samples with c.m. energies greater than 3.7~GeV. The red, green
and yellow histograms are the events in the sideband regions of
$f_{0}(980)$, $\etap$, and $\phi$, respectively.}
\label{fig:kpkmpippim_etap}
\end{figure}

Due to the low integrated luminosity and the relatively large
background level, the data sample at 3.686~GeV is not used to study $\EE\to\etapY$. After all the above event selections are applied, the distribution
of the $\phi f_{0}(980)$ invariant mass for the sum of data samples with
c.m.\ energies greater than 3.7~GeV is shown in
Fig.~\ref{fig:kpkmpippim_etap}, together with the distributions of
the events in $\etap$, $f_{0}(980)$ and $\phi$ sideband regions.
There are only a few events and no significant $\Y$
 or any other structure is observed.
Events from the sidebands can describe the events in the signal regions reasonably
well. The inclusive MC sample at 3.773~GeV is used to check the background
and no peaking background is found.

An unbinned maximum likelihood fit is applied to the sum of the $\phi
f_0(980)$ invariant mass distributions to determine the signal
yields for these data samples. We use the shape
from an exclusive MC simulation to describe the signal, and
use a second-order polynomial function for the background shape. The
resonant
parameters for the $\Y$ are taken from our measured values in the
previous study of $\EE\to \eta\Y$. The Bayesian method is used to estimate
the upper limit as described in Ref.~\cite{bes3_y4140}, and an
upper limit of 27.6 events is obtained at the 90\% C.L. after considering
the systematic uncertainty.

The upper limit on the ratio of the cross sections
$R=\sigma_{\etapY}/\sigma_{\etaY}$ is determined by
assuming this ratio is the same at different c.m.\ energy points, that is,
\begin{equation}
R=\frac{N^{\rm obs}_{\etapY}}{N^{\rm obs}_{\etaY}} \cdot
\frac{\mathcal{B}_{\eta}}{\mathcal{B}_{\etap}}\cdot \frac{\sum_i
\sigma_{\etaY}^i\cdot \mathcal{L}^{i}_{\rm int} \cdot
\epsilon_{\etaY}^{i} \cdot (1+\delta)^i \cdot (1+\delta^{\rm vac})^i}{\sum_j \sigma_{\etaY}^j\cdot
\mathcal{L}^{j}_{\rm int} \cdot \epsilon_{\etapY}^{j}\cdot (1+\delta)^j \cdot (1+\delta^{\rm vac})^j}.
\end{equation}
Here $N^{\rm obs}$ is the number of observed $\etaY$
($95.0\pm12.1$) or $\etapY$ events from the sum of the seven data samples;
 $\mathcal{B}_{\eta}$ and $\mathcal{B}_{\etap}$ are the
branching fractions of $\eta\to \gamma\gamma$ and $\etap\to
\gamma\pipi$~\cite{pdg}, respectively;
$\sigma_{\etaY}^i$ and $\mathcal{L}^{i}$ are the Born cross section
for $\EE\to\etaY$ and the integrated luminosity for the $i$-th data sample,
and the numbers are listed in Table~\ref{tab:cross_section};
$\epsilon^{i}$ is the reconstruction efficiency from MC simulation. With the numbers
obtained above, the upper limit on the ratio $R$ is estimated to
be 0.43 at the 90\% C.L., where the systematic uncertainties, which will be detailed later, are included.

\section{Systematic uncertainties}
\label{sec:sys}

\subsection{\boldmath Cross section measurement of $\EE\to\eta\Y$}
\label{sec:sys_crosssection}

Systematic uncertainties for the cross section measurement of $\EE\to \etaY$ are summarized in
Table~\ref{tab:sys_err} and are discussed below.

\begin{table*}[htbp]
\caption{Summary of systematic uncertainties (\%) in $\EE\to\eta\Y$ cross section
measurements for data samples at 8 different c.m. energies.}
\label{tab:sys_err}
\begin{tabular}{c c c c c c c c c}
\hline \hline
    Source/$\sqrt{s}$ (GeV) & 3.686 & 3.773 & 4.008 & 4.226 & 4.258 & 4.358 & 4.416 & 4.600 \\
    \hline
    Luminosity          & 1.0   & 1.0   & 1.0   & 1.0   & 1.0   & 1.0   & 1.0   & 1.0   \\
    Tracking            & 4.0   & 4.0   & 4.0   & 4.0   & 4.0   & 4.0   & 4.0   & 4.0   \\
    Photon              & 2.0   & 2.0   & 2.0   & 2.0   & 2.0   & 2.0   & 2.0   & 2.0   \\
    PID                 & 2.0   & 2.0   & 2.0   & 2.0   & 2.0   & 2.0   & 2.0   & 2.0   \\
    Branching fraction  & 1.2   & 1.2   & 1.2   & 1.2   & 1.2   & 1.2   & 1.2   & 1.2   \\
    Radiative correction& 2.6   & 4.2   & 2.4   & 6.5   & 5.1   & 2.7   & 2.6   & 3.7 \\
    Kinematic fit       & 0.4   & 0.3   & 0.4   &0.4    & 0.4   & 0.4   & 0.4   & 0.6 \\
    Background shape    & 57.6  & 14.7  & 84.3  & 4.8  & 9.2  & 11.4   & 45.9  & 5.5 \\
    Parameterization of $f_{0}(980)$ &
                          1.1  & 1.1 & 1.1 & 1.1 & 1.1 & 1.1 & 1.1 & 1.1 \\
    \hline
    Total               & 58.0  & 16.2  & 84.5  & 9.9  & 12.1  & 13.2  & 46.4  & 8.7 \\
    \hline \hline
\end{tabular}
\end{table*}

The luminosity is measured using large-angle Bhabha scattering with
an uncertainty less than 1.0\%~\cite{luminosity}. The difference
in detection efficiency between data and MC simulation in tracking
is 1.0\% per track, and that due to PID is taken
as 1.0\% per track~\cite{bes3_y4140}. The uncertainty in the
reconstruction efficiency for a photon is determined to be less than 1.0\% by studying a sample
of $J/\psi\to\rho\pi$ events.

The branching fractions of the $\eta$ and $\phi$ decays are taken
from the world average values in the PDG~\cite{pdg}, and the
corresponding uncertainties are taken as a systematic uncertainty. For
the $\eta$ and $\phi$ mass windows, the nominal values are taken to be $\pm 1.5\cdot$FWHM; the efficiency difference due
to any mass resolution difference between data and MC simulation is
very small and can be neglected compared to other sources of
uncertainties.

Since statistics are limited, the line shape of $\EE\to \etaY$
cannot be measured precisely. We assume there is no contribution
from charmonium(-like) states above 3.7~GeV and parameterize the
line shape to be proportional to $1/s^n$.
While we take the mean value of $n$ from a fit to the data by an iterative process, we vary $n$ by one standard deviation
and regenerate MC samples. The difference in $(1+\delta)\cdot\epsilon$
is taken as a systematic uncertainty.

To estimate the uncertainties introduced by the kinematic fit, we
use the same method described in Refs.~\cite{pull1,pull2}, $i.e.$,
correct the
track helix parameters in the MC sample and take half the
difference between results obtained with and without corrections
as systematic uncertainty (around 0.4\% for the all the data
samples). The MC sample with track parameter correction is
 used by default in the nominal analysis.

In the nominal fit, the shape from simulation of the non-$\Y$ process
$\EE\to\eta K^{+} K^{-} \pipi$ is taken to describe the
background. We change the shape of background to be a second order polynomial function for data with $\sqrt{s}>3.7$~GeV and to a shape from inclusive
MC sample at 3.686~GeV, and take the difference in signal yields
 as the systematic uncertainties. The uncertainty due to signal parametrization, which is obtained
    by altering the signal shape into a Breit-Winger function with a mass-dependent width, is found to
be negligible compared with that from the background shape. The systematic uncertainty
associated with the fit range is studied by changing fit range with 100 MeV/$c^2$,
the resultant value is 0.5\% only and is neglected.

The Flatt\'{e} formula~\cite{flatte} is used to model the
$f_{0}(980)$ lineshape in MC generation, where the parameters of $f_0(980)$ are from the BESII
experiment~\cite{flatte_bes2}. To estimate the corresponding
systematic uncertainty, we vary the parameters by one standard deviation
from the central values and the resultant difference in efficiency is
taken as the systematic uncertainty.

Assuming all the sources of uncertainty are independent,
the total uncertainty is obtained by summing all the individual uncertainties in quadrature, and is summarized in Table~\ref{tab:sys_err}.

\subsection{\boldmath Mass and width of the $\Y$}
\label{sec:sys_mass}

The systematic uncertainties for the mass and width of the $\Y$
include those from the mass calibration, signal shape of the $\Y$,
background shape and the c.m.\ energy.

A kinematic fit is performed with energy-momentum conservation, so
we can use the mass of $\eta$ to calibrate the mass of the $\Y$. A
simultaneous fit is performed on $M(\gamma\gamma)$ for all the
data samples. The difference between the fitted mass and the
nominal mass~\cite{pdg}, 2.1~MeV/$c^2$, is taken as the  systematic
uncertainty.

Since $\Y$ has $J^{PC}=1^{--}$, we expect it decays to $ \phi
f_{0}(980)$ in a relative $S$-wave. An $S$-wave Breit-Winger function
with mass-dependent width is used to parameterize the $\Y$ shape
in the fit, yielding a mass difference of 2.5~MeV/$c^2$ and a width
difference of 1.5~MeV. The mass resolution is about 4.5 MeV/$c^2$,
 which is much smaller than the width of $\Y$, and the corresponding effect on width
measurement is found
to be negligible.

In the nominal fit, we use the shape from simulated non-resonant
MC events to describe the background. To study the corresponding
systematic uncertainty, we change the background shape to a
second-order polynomial function and the resultant
 differences in fitted mass and
width, 8.2~MeV/$c^2$ and 12.1~MeV, respectively, are taken as systematic uncertainties.

The c.m.\ energy of the $\EE$ system also affects the determination of the mass and
width of the $\Y$ due to the kinematic constraint between initial
and final states. An analysis~\cite{cmsenergy} reveals that the
uncertainty on c.m.\ energy of $\EE$ is less than 0.6~MeV. We
change the c.m.\ energy by $\pm$0.6~MeV in the kinematic fit and
study the changes of mass and width, which are 0.2~MeV/$c^2$ and
0.4~MeV, respectively.

The quadratic sum of all the above uncertainties, 8.8~MeV/$c^2$ and 12.2~MeV
for the mass and width, respectively, are taken as the total
uncertainties.

\subsection{\boldmath Branching fraction $\mathcal{B}(\psip\to \etaY)$}
\label{sec:sys_psip}

The sources of systematic uncertainties on the product of
branching fractions $\mathcal{B}(\psip\to \etaY)
\cdot\mathcal{B}(\Y\to \phi f_{0}(980) \to \phi \pipi)$ are the
same as those in the cross section measurement.
An additional uncertainty associated with the total number of $\psip$ events
~\cite{npsip_2}, 0.66\%, is also taken into account. The resultant systematic uncertainty
for the branching fraction $\mathcal{B}(\psip\to \etaY)$ is 57.8\%.

\subsection{\boldmath Ratio $R=\sigma(\EE\to \etapY)/\sigma(\EE\to \etaY)$}
\label{sec:sys_etap}

For the ratio $R$, the common systematic uncertainties between $\EE\to
\eta\Y$ and $\etapY$ cancel and the remaining
uncertainties arise from the differences between $\eta$ and
$\etap$ reconstruction, where $\eta$ is reconstructed from two
photons and $\etap$ from one photon and two charged pions. The
fraction of common systematic uncertainty introduced by the
kinematic fit is hard to estimate. To be conservative, we assume
they are independent and the quadratic sum of them is taken as the
uncertainty of $R$. The systematic uncertainty due to the background
shape, 48.7\%, is obtained by varying the shape to that determined by the events in sideband regions of $\etap$ and
$\phi$. We assume that all the sources of systematic uncertainty are independent and obtain
the total uncertainty in $R$ as a quadratic sum of statistical and
systematic uncertainties, which is 50.6\%, and is considered in
calculating the upper limit of $R$.

\section{Summary}
\label{sec:summary}

We observe clear $\Y$ signals in the process $\EE\to \etaY$ using
data samples at $\sqrt{s}=3.773$, 4.008, 4.226, 4.258, 4.358,
4.416, and 4.600~GeV. In the measured c.m.\ energy dependent
Born cross sections, no obvious peaks corresponding to decays of charmonium(-like) states to the final state $\etaY$ are
seen. The mass and width of the $\Y$ are
measured to be ($2135\pm 8\pm 9$)~MeV/$c^2$ and ($104\pm 24\pm 12$)~MeV,
respectively, where the first uncertainties are statistical and
the second systematic. The results are consistent with previous
measurements~\cite{babar_1,belle,bes2_y,babar_2,bes3_y}, and the
width tends to be larger but similar with the results of Belle and
BESIII~\cite{belle,bes3_y}. An examination of the Dalitz plot of
the $\Y\to \phi\pipi$ indicates that $\phi f_0(980)$ is a dominant component, and no obvious signal of a potential charged strangeonium-like state
$Z_{s}\to \phi\pi$ is observed.

The cross section of $\EE\to \etaY$ varies with c.m.\ energy as
 $1/s^n$ with $n=2.65\pm0.86$, which can be compared with
measurements of other vector-pseudoscalar final states and
theoretical calculations~\cite{pqcd_vp,belle_VP}. The deviation from the
behavior of final states with ordinary vector quarkonium states
may reveal the nature of the $\Y$, where theoretical calculations
are expected for different assumptions of the parton configuration
of the $\Y$.

No significant $\psip\to \etaY)$ signal is observed, and the
product branching fraction $\mathcal{B}(\psip\to \etaY)
\cdot\mathcal{B}(\Y\to \phi f_{0}(980)\to \phi \pipi)$ is obtained to be $(0.81\pm
0.97\pm 0.47)\times 10^{-6}$, or less
than $2.2\times 10^{-6}$ at the 90\% C.L. The ratio of the
branching fractions $\mathcal{B^*}(\psip\to
\etaY)/\mathcal{B^*}(J/\psi\to \etaY)$ is $(0.23\pm 0.29\pm 0.13)$\%,
 or
less than 0.65\% at the 90\% C.L.,
after considering the phase space difference between $\psip$ and
$\jpsi$ decays to the $\eta\Y$ final state. A large suppression of $\psip\to \etaY$ with
respect to the 12\% rule~\cite{rule12} is observed. This is therefore
another vector-pseudoscalar channel failing the 12\% rule.

With the same data samples, the process $\EE\to \etapY$ is
searched for. No significant signal is observed. We set the
upper limit on the ratio of the cross sections $\sigma(\EE\to
\etapY)/\sigma(\EE\to \etaY)<0.43$ at the 90\% C.L.


\section*{Acknowledgement}
The BESIII collaboration thanks the staff of BEPCII and the IHEP computing center for their strong support. This work is supported in part by National Key Basic Research Program of China under Contract No. 2015CB856700; National Natural Science Foundation of China (NSFC) under Contracts Nos. 10979033, 11235011, 11335008, 11425524, 11625523, 11635010; the Chinese Academy of Sciences (CAS) Large-Scale Scientific Facility Program; the CAS Center for Excellence in Particle Physics (CCEPP); Joint Large-Scale Scientific Facility Funds of the NSFC and CAS under Contracts Nos. U1332201, U1532257, U1532258; CAS under Contracts Nos. KJCX2-YW-N29, KJCX2-YW-N45, QYZDJ-SSW-SLH003; 100 Talents Program of CAS; New Century Excellent Talents
in University (NCET) under Contract No. NCET-13-0342; Shandong Natural
Science Funds for Distinguished Young Scholar under Contact No. JQ201402; National 1000 Talents Program of China; INPAC and Shanghai Key Laboratory for Particle Physics and Cosmology; German Research Foundation DFG under Contracts Nos. Collaborative Research Center CRC 1044, FOR 2359; Istituto Nazionale di Fisica Nucleare, Italy; Koninklijke Nederlandse Akademie van Wetenschappen (KNAW) under Contract No. 530-4CDP03; Ministry of Development of Turkey under Contract No. DPT2006K-120470; National Science and Technology fund; The Swedish Research Council; U. S. Department of Energy under Contracts Nos. DE-FG02-05ER41374, DE-SC-0010118, DE-SC-0010504, DE-SC-0012069; University of Groningen (RuG) and the Helmholtzzentrum fuer Schwerionenforschung GmbH (GSI), Darmstadt; WCU Program of National Research Foundation of Korea under Contract No. R32-2008-000-10155-0.


\end{document}

%% file: authors_May2016.tex
\author{
\small
M.~Ablikim$^{1}$, M.~N.~Achasov$^{9,d}$, S. ~Ahmed$^{14}$, X.~C.~Ai$^{1}$, O.~Albayrak$^{5}$, M.~Albrecht$^{4}$, D.~J.~Ambrose$^{45}$, A.~Amoroso$^{50A,50C}$, F.~F.~An$^{1}$, Q.~An$^{47,38}$
% ...
\\
\vspace{0.2cm}
(BESIII Collaboration)\\
\vspace{0.2cm}
{\it $^{1}$ Institute of High Energy Physics, Beijing 100049, People's Republic of China\\
% ...
$^{4}$ Bochum Ruhr-University, D-44780 Bochum, Germany\\
$^{5}$ Carnegie Mellon University, Pittsburgh, Pennsylvania 15213, USA\\
% ...
$^{9}$ G.I. Budker Institute of Nuclear Physics SB RAS (BINP), Novosibirsk 630090, Russia\\
% ...
$^{14}$ Helmholtz Institute Mainz, Johann-Joachim-Becher-Weg 45, D-55099 Mainz, Germany\\
% ...
$^{38}$ State Key Laboratory of Particle Detection and Electronics, Beijing 100049, Hefei 230026, People's Republic of China\\
% ...
$^{45}$ University of Rochester, Rochester, New York 14627, USA\\
% ...
$^{47}$ University of Science and Technology of China, Hefei 230026, People's Republic of China\\
% ...
$^{50}$ (A)University of Turin, I-10125, Turin, Italy; (B)University of Eastern Piedmont, I-15121, Alessandria, Italy; (C)INFN, I-10125, Turin, Italy\\
% ...
$^{d}$ Also at the Novosibirsk State University, Novosibirsk, 630090, Russia\\
% ...
}
}

\affiliation{}